\documentclass[]{spie}  
\usepackage[]{graphicx}

\title{Photonic spin Hall effect for precision metrology}

\author{Xinxing Zhou, Shizhen Chen, Yachao Liu, Hailu Luo$^{*}$, and Shuangchun Wen$^{\dagger}$
\skiplinehalf \supit{}Laboratory for Spin Photonics, College of
Physics and Microelectronic Science, Hunan University, Changsha
410082, China\nonumber}

\authorinfo{\\$^{*}$Hailu Luo email address: hailuluo@hnu.edu.cn\\$^{\dagger}$Shuangchun Wen email address:
scwen@hnu.edu.cn}

\begin{document}
  \maketitle

\begin{abstract}
The photonic spin Hall effect (SHE) is generally believed to be a
result of an effective spin-orbit coupling, which describes the
mutual influence of the spin (polarization) and the trajectory of
the light beam. The photonic SHE holds great potential for precision
metrology owing to the fact that the spin-dependent splitting in
photonic SHE are sensitive to the physical parameter variations of
different systems. Remarkably, using the weak measurements, this
tiny spin-dependent shifts can be detected with the desirable
accuracy so that the corresponding physical parameters can be
determined. Here, we will review some of our works on using photonic
SHE for precision metrology, such as measuring the thickness of
nanometal film, identifying the graphene layers, detecting the
strength of axion coupling in topological insulators, and
determining the magneto-optical constant of magnetic film.
\end{abstract}


\keywords{photonic spin Hall effect, spin-orbit coupling, weak
measurements}

\section{INTRODUCTION}
The photonic spin Hall effect (SHE) manifests itself as
spin-dependent splitting of left- and right-handed circularly
polarized components when a spatially confined light beam is
reflected or transmitted at an
interface~\cite{Onoda2004,Bliokh2006,Hosten2008}. The photonic SHE
can be regarded as a direct optical analogy of SHE in an electronic
system, in which the spin photons play the role of the spin charges
and a refractive index gradient plays the role of the applied
electric field. This interesting phenomenon is generally believed to
be a result of an effective spin-orbit coupling, which describes the
mutual influence of the spin (polarization) and the trajectory of
the light beam. There are two types of geometric phases playing the
important role in photonic SHE: the spin redirection Berry phase and
the Pancharatnam-Berry phase~\cite{Bliokh2008a,Bliokh2008b}. The
photonic SHE is currently attracting growing attention and has been
intensively investigated in different physical systems such as
optical
physics~\cite{Aiello2008,Luo2009,Qin2009,Hermosa2011,Luo2011a,Ling2014},
high-energy physics~\cite{Gosselin2007,Dartora2011}, semiconductor
physics~\cite{Menard2009,Menard2010}, and
plasmonics~\cite{Shitrit2011,Gorodetski2012,Shitrit2013,Yin2013,Kapitanova2014}.

Remarkably, the spin-dependent splitting in photonic SHE are
sensitive to the physical parameter variations of different systems,
and therefore it holds great potential applications in precision
metrology. However, the spin-dependent splitting of photonic SHE in
these systems is just a few tens of nanometers so that the actual
equipment can not distinguish it directly. We resolve this problem
by using the precise signal enhancement technique called quantum
weak measurements which has attracted a lot of
attention~\cite{Aharonov1988,Ritchie1991,Dixon2009,Lundeen2011,Kocsis2011,Dressel2014,Jordan2014,Knee2014,Zhou2014}.
The idea of weak measurements can be described as follows: if we
initially select the quantum system with a well-defined preselection
state, the corresponding large expectation values can be obtained
with a suitable postselection state, which makes the eigenvalues to
be clearly distinguished~\cite{Jozsa2007,Dennis2012,Lorenzo2014}.
Using the quantum weak measurement, the photonic SHE can be detected
with the desirable accuracy.

In this paper, we will review some of our recent works on using
photonic SHE for precision metrology, such as measuring the
thickness of nanometal film~\cite{Zhou2012a}, identifying the
graphene layers~\cite{Zhou2012b}, detecting the strength of axion
coupling in topological insulators~\cite{Zhou2013}, and determining
the magneto-optical constant of magnetic film. We find that the
physical parameter variations in these systems can effectively
change the spin-dependent displacements. We firstly establish the
quantitative relationship between the spin-dependent shifts and the
physical parameters. After detecting the spin-dependent
displacements with weak measurement method, we can accurately
determine these physical parameters. The rest of the paper is
organized as follows. In Sec. 2, we establish a general propagation
model to describe the photonic SHE on the sample. In the Sec. 3, we
firstly introduce the weak measurements experimental process. Then,
we will briefly review our recent works on using photonic SHE for
precision metrology. Finally, a conclusion is given in Sec. 4.

\begin{figure}
\centerline{\includegraphics[width=8.5cm]{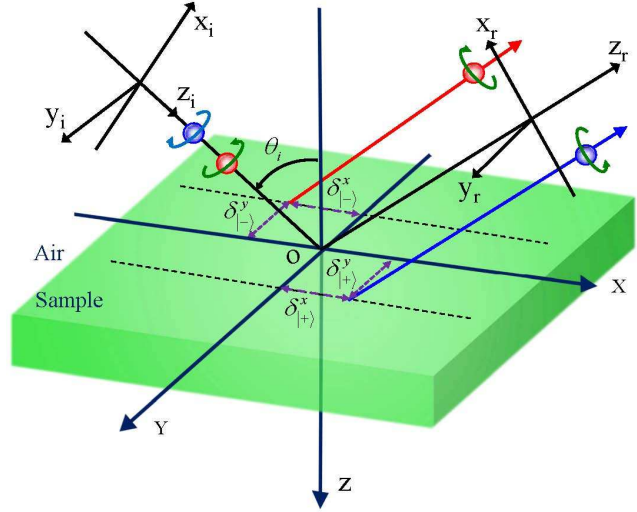}}
\caption{\label{Fig1} Schematic of photonic SHE on a sample. The
sample can be nanometal film, graphene film, topological insulators,
and magnetic film. A linearly polarized beam reflects on the sample
and then splits into left- and right-handed circularly polarized
components, respectively. $\delta^{x}_{|+\rangle}$ and
$\delta^{x}_{|-\rangle}$ indicate the in-plane shift of left- and
right-handed circularly polarized components.
$\delta^{y}_{|+\rangle}$ and $\delta^{y}_{|-\rangle}$ denote the
transverse spin-dependent displacements. Here, $\theta_{i}$ is the
incident angle. }
\end{figure}

\section{GENERAL PROPAGATION MODEL}
Figure~\ref{Fig1} schematically draws the photonic SHE of light beam
reflection from a sample interface. We firstly establish the
quantitative relationship between the spin-dependent shifts in
photonic SHE and the physical parameters of sample. Here, the
physical parameters include the thickness of nanometal film,
graphene's layers, axion angle in topological insulators, and the
magneto-optical constant of magnetic film. The incident polarization
states are chosen as $|H\rangle$ and $|V\rangle$. In the spin basis,
the horizontal and vertical polarization states can be expressed as
$|H\rangle=(|+\rangle+|-\rangle)/{\sqrt{2}}$ and
$|V\rangle=i(|-\rangle-|+\rangle)/{\sqrt{2}}$. Corresponding, the
states of reflected beam can be obtained:
\begin{equation}
|H\rangle\rightarrow\frac{r_{p}}{\sqrt{2}}\left[\exp(+ik_{ry}\delta^{H}_{r})|+\rangle+\exp(-ik_{ry}\delta^{H}_{r})|-\rangle\right]\label{H
spectrum},
\end{equation}
\begin{equation}
|V\rangle\rightarrow\frac{ir_{s}}{\sqrt{2}}\left[-\exp(+ik_{ry}\delta^{V}_{r})|+\rangle+\exp(-ik_{ry}\delta^{V}_{r})|-\rangle\right]\label{V
spectrum}.
\end{equation}
In the above equations,
$\delta^{H}_{r}=(1+r_{s}/r_{p})\cot\theta_{i}/k_{0}$,
$\delta^{V}_{r}=(1-r_{p}/r_{s})\cot\theta_{i}/k_{0}$. We should note
that, as for the topological insulators, the states of reflected
beam have different forms. The detailed discussions can be found in
our previous work~\cite{Zhou2013}.

The photonic SHE manifests for the spin-dependent splitting of left-
and right-handed circularly polarized components. We consider the
spin separation in the x direction (in-plane shift) and y direction
(transverse shift). In the following, we calculate the shifts of
these two spin components. The wavefunction of reflected photons is
composed of the packet spatial extent $\phi(k_{ry})$ and the
polarization description $|H,V\rangle$:
\begin{equation}
|\Phi^{H,V}\rangle=\int dk_{ry}
\phi(k_{ry})|k_{ry}\rangle|H,V\rangle\label{inital}.
\end{equation}
After photons reflection from the sample interface, the initial
state $|\Phi_{inital}^{H,V}\rangle$ evolve into the final state
$|\Phi_{final}^{H,V}\rangle$. As a result of spin-orbit coupling,
the shifts of the two spin components compared to the
geometrical-optics prediction are given by
\begin{equation}
\delta_{|\pm\rangle}^{H,V}=\frac{\langle
\Phi^{H,V}|i\partial_{\mathbf{k_\perp}}|{\Phi^{H,V}}\rangle}{\langle
\Phi^{H,V}|\Phi^{H,V}\rangle}.\label{BCII}
\end{equation}
Here, we suppose the $\phi(k_{ry})$ is a Gaussian wave function.
Calculating the reflected shifts of photonic SHE requires the
explicit solution of the boundary conditions at the sample
interfaces. As for the nanometal film and graphene film, we need to
deal with the multilayer structure model. Thus, we need to know the
generalized Fresnel reflection of the sample,
\begin{eqnarray}
r_{A}=\frac{R_{A}+R_{A}^{'}\exp(2ik_{0}\sqrt{n^{2}-\sin^{2}\theta_{i}}d)}{1+R_{A}R_{A}^{'}\exp(2ik_{0}\sqrt{n^{2}-\sin^{2}\theta_{i}}d)}.
\end{eqnarray}
Here, $A\in\{p,s\}$, $R_{A}$ and $R_{A}^{'}$is the Fresnel
reflection coefficients at the first interface and second interface,
respectively. $n$ and $d$ represent the refractive index and the
thickness of the nanometal film and graphene film, respectively.

\begin{figure}
\centerline{\includegraphics[width=10cm]{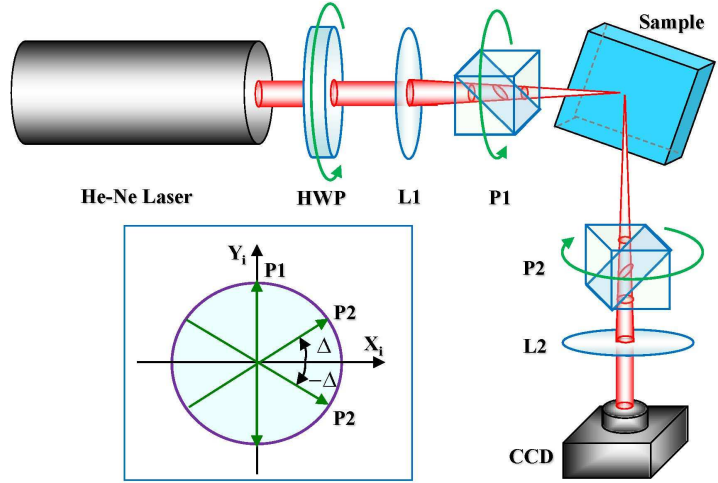}}
\caption{\label{Fig2} The experimental setup in weak measurements.
The sample is a BK7 prism prepared with the nanometal film, graphene
film, topological insulators, and magnetic film. L1 and L2, lenses
with effective focal length $50\mathrm{mm}$ and $250\mathrm{mm}$,
respectively. HWP, half-wave plate (for adjusting the intensity). P1
and P2, Glan Laser polarizers. CCD, charge-coupled device (Coherent
LaserCam HR). The light source is a $21\mathrm{mW}$ linearly
polarized He-Ne laser at $632.8\mathrm{nm}$ (Thorlabs HNL210L-EC).
The inset shows the states of preselection and postselection. Here,
the preselection state is prepared in $|V\rangle$. }
\end{figure}

\section{PHOTONIC SPIN HALL EFFECT FOR PRECISION METROLOGY}

We have established the relationship between the physical parameters
of sample and the spin-dependent displacements induced by photonic
SHE. Next, we will use the weak measurements method to detect the
this tiny shifts. After detecting the spin-dependent displacements,
we can accurately determine these physical parameters. As shown in
Fig.~\ref{Fig2}, our experimental setup is similar to that in
Ref~\cite{Luo2011}. Our samples are the usual BK7 prism prepared
with the nanometal film, graphene film, topological insulators, and
magnetic film. A Gauss beam generated by He-Ne laser is firstly
focused by the lens (L1) and experiences preselection in the state
$|\psi_{1}\rangle$=$|H\rangle$ or $|V\rangle$ with the polarizer P1.
When the light beam reflects from the sample interface, the photonic
SHE happens allowing for the left- and right-handed circularly
polarized components splitting in the x and y directions
corresponding to the in-plane and transverse displacements. This
process can be seen as the weak interaction allowing for the
coupling between the observable and the meter. And then the beam
passes through the second polarizer P2 preparing for the
postselection state $|\psi_{2}\rangle=|V\pm\Delta\rangle$ or
$|H\pm\Delta\rangle$. At the surface of the second polarizer, the
two spin components experience destructive interference making the
enhanced shift in the meter much larger than the initial one.
Calculating the reflected field distribution yields the amplified
shifts of photonic SHE. After passing through the second lens (L2),
a CCD is used to capture the optical signal and measure the
amplified shifts. The process discussed above is called the weak
value amplification and $\Delta$ is the postselection angle. We
should note that the imaginary weak value also corresponds to a
shift of the meter in momentum space, which leads to the possibility
of even larger enhancements following the beam free evolution. This
process can be seen as propagation amplification that produces the
amplified factor F. In the following, we will review some of our
recent works on using photonic SHE for precision metrology.

\begin{figure}
\centerline{\includegraphics[width=10cm]{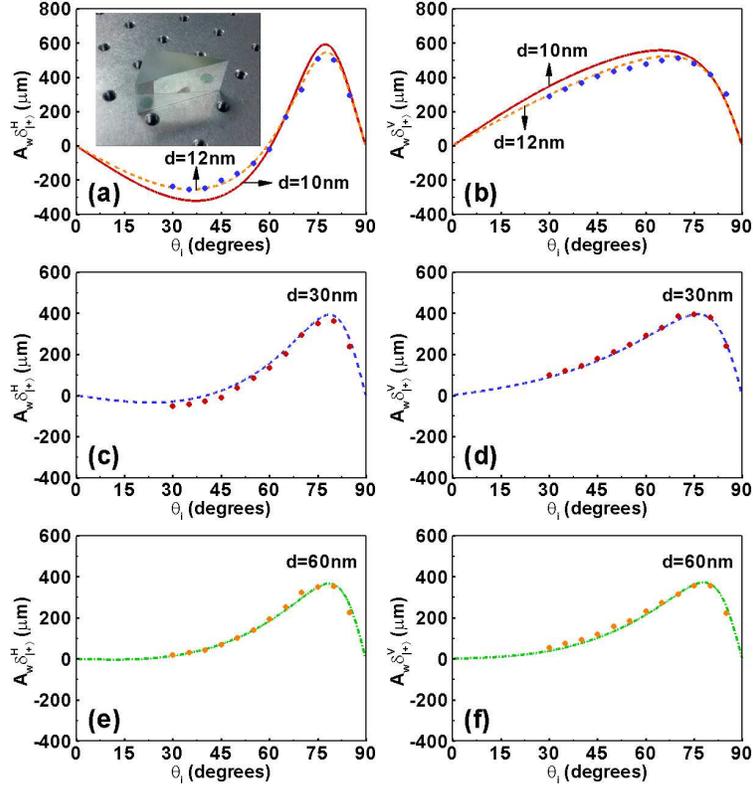}}
\caption{\label{Fig3} In the case of horizontal polarized (left
column) and vertical polarized (right column), the amplified
displacements of light beam reflection on Ag film with different
thicknesses: [(a),(b)] $10\mathrm{nm}$ and $12\mathrm{nm}$,
[(c),(d)] $30\mathrm{nm}$, and [(e),(f)] $60\mathrm{nm}$. The lines
show the theoretical value and the dots denote the experimental
results. The inset shows the experimental sample: a BK7 prism coated
with Ag film.~\cite{Zhou2012a} }
\end{figure}

In the fist work, we have used the photonic SHE to measure the
thickness of nanometal film~\cite{Zhou2012a}. We establish a general
propagation model to describe the photonic SHE on a nanometal film
and reveal the impact of the corresponding physical parameters on
the spin-dependent splitting in photonic SHE. It is well known that
the photonic SHE manifests itself as the spin-orbit coupling. We
find that the spin-orbit coupling in the photonic SHE can be
effectively modulated by adjusting the thickness of the metal film.
A similar effect can also be observed in layered nanostructures, in
which the transverse displacement changes periodically with the air
gap increasing or decreasing. Additionally, the transverse
displacement is sensitive to the thickness of metal film in certain
range for horizontal polarization light beam. We also note tha a
large negative transverse shift can be observed.

Next, we focus our attention on the weak measurements experiment.
Here, the BK7 glass substrate coated Ag film is chosen as our sample
(with three different thickness $10\mathrm{nm}$, $30\mathrm{nm}$ and
$60\mathrm{nm}$). The experimental setup is described in the above
contents. We measure the displacements of photonic SHE on the
nanometal film every $5^{\circ}$ from $30^{\circ}$ to $85^{\circ}$
in the case of horizontal and vertical polarization, respectively.
Limited by the large holders of the lens, polarizers and He-Ne
laser, displacements at small incident angles were not measured. It
should be noted that the experimental results are in good agreement
with the theoretical ones when the film thicknesses are
$30\mathrm{nm}$ and $60\mathrm{nm}$. However, we observe a small
deviation when the thickness is $10\mathrm{nm}$. Note that the
thickness of the nanometal film has an error limited by the
experimental condition. When the thickness reaches to
$10\mathrm{nm}$, the SHE of light is very sensitive to the error. It
is the reason why there is a small deviation between the
experimental and the theoretical data. From the experimental
results, we can conclude that the actual thickness of the film is
about $12\mathrm{nm}$ (Fig.~\ref{Fig3}). These findings provide a
pathway for modulating the photonic SHE and thereby open the
possibility of developing nanophotonic applications.

We also propose using the photonic SHE to identify the graphene
layers~\cite{Zhou2012b}. The quick and convenient technique for
identifying the layer numbers of graphene film is important for
accelerating the study and exploration of graphene material. There
have many methods for determining the layer numbers of graphene
film, yet existing limitation. For example, atomic force microscopy
technique is the straight way to determine the layer numbers of
graphene. But this method shows a slow throughput and may induce
damage to the sample. Unconventional quantum Hall
effects~\cite{Zhang2005} are usually used to distinguish one layer
and two layers graphene from multiple layers. Raman
spectroscopy~\cite{Gupta2006} shows characteristic for quick and
nondestructive measuring the layer numbers of graphene. However, it
is not obvious to tell the differences between bilayer and a few
layers of graphene films~\cite{Ni2007}. We find that the photonic
SHE can serve as a useful metrological tool for characterizing the
structure parameters' variations of nanostructure due to their
sensitive dependence. So, the photonic SHE may have a potential to
determine the layer numbers of graphene.

\begin{figure}
\centerline{\includegraphics[width=17cm]{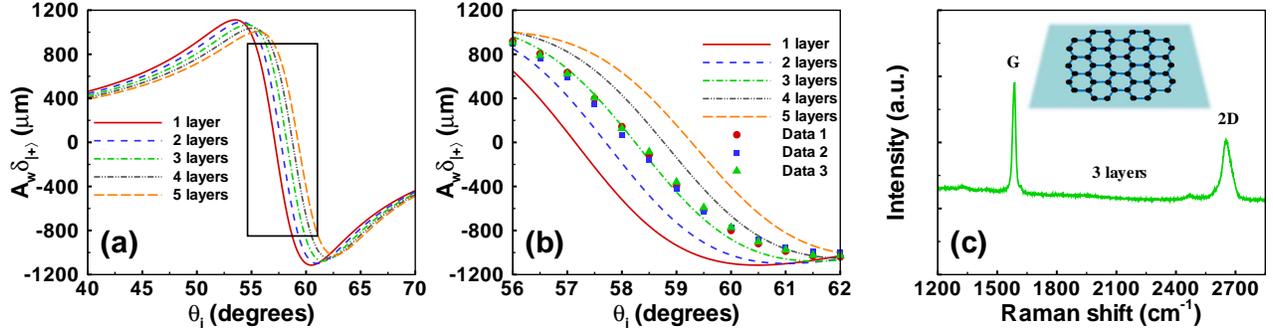}}
\caption{\label{Fig4} Experimental results for determining the layer
numbers of graphene. (a) shows the theoretical spin-dependent shifts
in the case of graphene layer numbers changing from one to five. (b)
describes the experimental results for determining the layer numbers
of graphene. The lines represent the theoretical results. The
circle, square and triangle show the experimental data obtained from
three different areas of the graphene sample. (c) Raman spectroscopy
of the sample. The inset shows the graphene sample.~\cite{Zhou2012b}
}
\end{figure}

We establish the relationship between the spin-dependent
displacements and the graphene layer numbers. The weak measurements
method has been used to detect the transverse shifts, and so the
graphene layer numbers can be obtained. However, there exists two
unknown parameters (refractive index and layer numbers of graphene)
to be identified. Before identifying the graphene layers, we need to
choose the suitable refractive index parameter of graphene. We
choose one suitable refractive index according from the work of
Bruna and Borini~\cite{Bruna2009}. Here, the refractive index of
graphene is about $3.0+1.149i$ at 633 nm. Through measuring the
spin-dependent shifts of photonic SHE on the graphene film, we prove
that this refractive index is suitable for actual situation. Using
the suitable refractive index n=$3.0+1.149i$ at 633 nm, we can
identify the layer numbers of an unknown graphene film. It should be
noted that we cannot fabricate the graphene film with the precise
layer numbers when the graphene film has more than two layers. We
just know the approximate layer numbers ranging from three to five
layers. We want to determine the actual layer numbers of this
graphene film. As shown in Fig.~\ref{Fig4}, we measure the
transverse displacements with the incident angle changing from
$56^{\circ}$ to $62^{\circ}$. To avoid the influence of impurities
and other surface quality factors of graphene film, we carried out
the experiments for three different areas of the graphene sample. It
is concluded that the actual layer numbers of the film is three.

\begin{figure}
\centerline{\includegraphics[width=10cm]{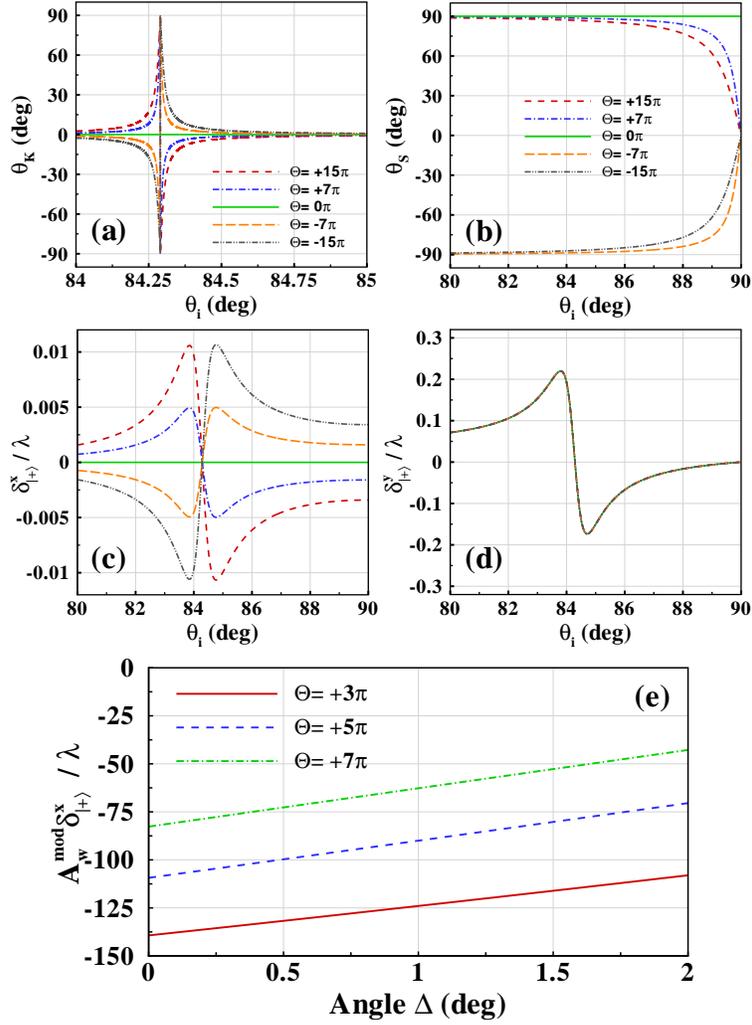}}
\caption{\label{Fig5} The photonic SHE and magneto-optical Kerr
effect induced by axion coupling at air-TI interfaces in the case of
horizontal polarization (a-d) and the corresponding weak
measurements (e). Here the parameters are the refractive index of
TIs n=10 (appropriate for the TIs such as $Bi_{1-x}Se_{x}$). The
beam waist is selected as $w_{0}=20\lambda$. In the weak
measurements process, the incident angle is chosen as
$\theta_{i}=84^{\circ}$.~\cite{Zhou2013} }
\end{figure}

Recently, the topological insulators (TIs) material has aroused
tremendous interest~\cite{Qi2010,Moore2009}. It has gapless helical
surface states owing to the topological protection of the
time-reversal symmetry and represents a full energy gap in the
bulk~\cite{Fu2007,Maciejko2010}. In a recent paper, we theoretically
investigate the photonic SHE of a Gaussian beam reflected from the
interface between air and topological insulators
(TIs)~\cite{Zhou2013}. We reveal that the spin-orbit coupling effect
in TIs can be routed by adjusting the axion angle variations. It is
shown that the magneto-optical Kerr effect can be significantly
altered due to the axion coupling and shows close relationship with
spin-dependent splitting in photonic SHE [Fig.~\ref{Fig5}(a)
and~\ref{Fig5}(b)]. We find that, unlike the transverse
spin-dependent splitting, the in-plane one is sensitive to the axion
angle [Fig.~\ref{Fig5}(c) and~\ref{Fig5}(d)]. Due to the the
limitation of experimental condition, we theoretically propose a
weak measurement method to determine the strength of axion coupling
by probing the in-plane splitting of the photonic SHE.

\begin{figure}
\centerline{\includegraphics[width=10cm]{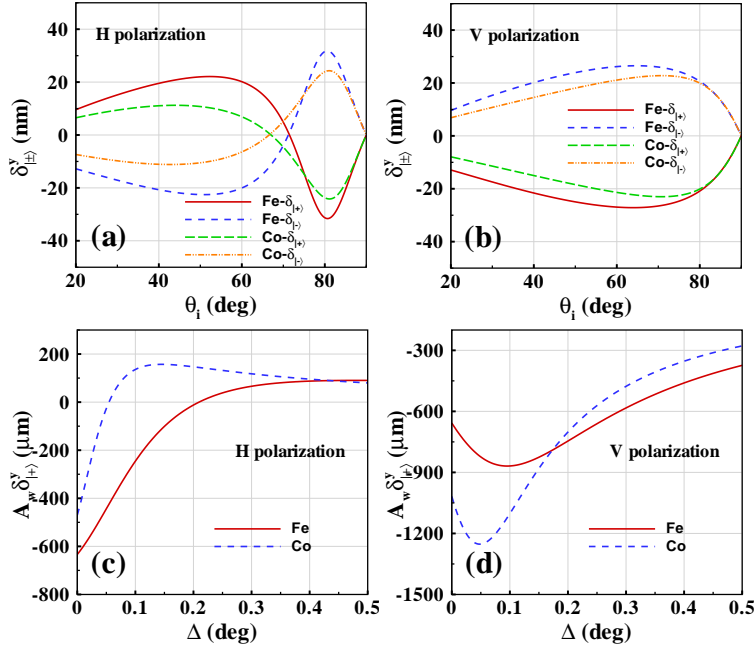}}
\caption{\label{Fig6} The preliminary results for determining the
magneto-optical constants. (a) and (b) describe the initial
spin-dependent shifts of Fe and Co materials in the case of H and V
polarizations. The corresponding amplified displacements under the
condition of H and V polarizations are shown in (c) and (d). }
\end{figure}

The incident beam is focused by the lens and is preselected in the
horizontal polarization, and then it is postselected in the
polarization state with $|V\pm\Delta\rangle$ ($\Delta$ is the
amplified angle). The relevant amplitude of the reflected field at a
plane can be obtained, allowing for calculation of amplified
displacement. The theoretical amplified shifts are shownin
Fig.~\ref{Fig5}(e). Here the incident angle is fixed to
$84^{\circ}$. Then we obtain the amplified displacements varying
with axion angle and amplified angle. For a fixed angle $\Delta$,
the amplified in-plane shifts change clearly with the different
axion angles, and so we can measure the axion coupling effect by
determining the in-plane displacements with weak measurements. These
findings offer us potential methods for determining the strength of
the axion coupling and provide new insight into the interaction of
light with TIs.

We have also used the photonic SHE for determining the
magneto-optical constant of magnetic film and the preliminary
results can be seen in Fig.~\ref{Fig6}. The magneto-optical constant
is an important parameter for the study and exploration of magnetic
material. The relationship between the spin-dependent splitting in
photonic SHE and the magneto-optical constant of magnetic film is
established. Here, we choose the Fe and Co as our samples, which
have different magneto-optical constants
($Q_{Fe}$=$0.0215-0.0016i$~\cite{Johnson1974,Yang1993} and
$Q_{Co}$=$0.0189-0.0043i$~\cite{Osgood1997} at 633 nm). From the
Fig.~\ref{Fig6}(a) and~\ref{Fig6}(b), we can see that the
spin-dependent displacements are sensitive to the magneto-optical
constants of different magnetic materials. So, we can determine the
magneto-optical constants by measuring the spin-dependent splitting
of photonic SHE. In our experiment, the weak measurements has been
used to detect this tiny shifts [the preliminary theoretical results
are shown in Fig.~\ref{Fig6}(c) and~\ref{Fig6}(d)]. We also find
that the amplified spin shifts are sensitive to the variations of
magneto-optical constants (Fe and Co). Importantly, the Kerr
rotation angle in magneto-optical Kerr effect can also be detected
by using this way, which shows higher accuracy than the normal
extinction method. In fact, our experiment is in progress and the
manuscript is in preparation.

\section{CONCLUSIONS}
In summary, we have reviewed our recent works on using photonic SHE
for precision metrology. After establishing the quantitative
relationship between the spin-dependent shifts in photonic SHE and
the physical parameters in different systems, we have used the weak
measurements methods for measuring the thickness of nanometal film,
identifying the graphene layers, detecting the strength of axion
coupling in topological insulators, and determining the
magneto-optical constant of magnetic film. These findings provide a
practical application for photonic SHE and thereby open the
possibility of developing spin-based nanophotonic device.

\acknowledgments  
This research was partially supported by the National Natural
Science Foundation of China (Grants Nos. 61025024 and 11274106) and
Hunan Provincial Innovation Foundation for Postgraduate (Grant No.
CX2013B130).

\end{document}